\documentclass[sigconf,natbib=true,nonacm]{acmart}

\usepackage{graphicx} 

\usepackage{multirow}
\usepackage{bbding} 
\usepackage{float}
\floatstyle{plaintop}
\restylefloat{table}
\usepackage{subfigure}
\usepackage{amsthm}
\newtheorem{lemma}{Lemma}[section]
\newtheorem{assumption}{Assumption}[section]

\usepackage{amsmath}
\DeclareMathOperator*{\argmin}{arg\,min}
\DeclareMathOperator*{\argmax}{arg\,max}
\usepackage{bm}

\usepackage{colortbl}
\usepackage[normalem]{ulem}
\useunder{\uline}{\ul}{}
\usepackage{xcolor}
\usepackage{color}  
\definecolor{shadecolor}{rgb}{239,239,239}  
\usepackage{framed}
\usepackage{pifont}
\usepackage{multirow}

\usepackage[capitalize]{cleveref}
\crefname{section}{Sec.}{Secs.}
\Crefname{section}{Section}{Sections}
\Crefname{table}{Table}{Tables}
\crefname{table}{Tab.}{Tabs.}
\crefname{algorithm}{Algo.}{Algos.}

\def\dataset{MSDR}
\def\model{DSFNet}

\newcommand{\yes}{\ding{52}}
\newcommand{\no}{\ding{56}}

\AtBeginDocument{%
  \providecommand\BibTeX{{%
    \normalfont B\kern-0.5em{\scshape i\kern-0.25em b}\kern-0.8em\TeX}}}

\setcopyright{acmlicensed}
\copyrightyear{2018}
\acmYear{2018}
\acmDOI{XXXXXXX.XXXXXXX}

\acmConference[Conference acronym 'XX]{Make sure to enter the correct
  conference title from your rights confirmation emai}{June 03--05,
  2018}{Woodstock, NY}
\acmISBN{978-1-4503-XXXX-X/18/06}

\begin{document}

\title{DSFNet: Learning Disentangled Scenario Factorization for Multi-Scenario Route Ranking}


\author{Jiahao Yu}
\authornote{These authors contributed equally to this research. Work done when Jiahao Yu was an intern at Alibaba Group. Longfei Xu is the corresponding author.}
\affiliation{%
  \institution{School of Software, BNRist, Tsinghua University}
  \city{Beijing}
  \country{China}}
\email{yujh21@mails.tsinghua.edu.cn}

\author{Yihai Duan}
\authornotemark[1]
\affiliation{%
  \institution{Alibaba Group}
  \city{Beijing}
  \country{China}}
\email{duanyihai.dyh@alibaba-inc.com}

\author{Longfei Xu}
\authornotemark[1]
\affiliation{%
  \institution{Alibaba Group}
  \city{Beijing}
  \country{China}}
\email{longfei.xl@alibaba-inc.com}

\author{Chao Chen}
\affiliation{%
  \institution{Alibaba Group}
  \city{Beijing}
  \country{China}}
\email{cc201598@alibaba-inc.com}

\author{Shuliang Liu}
\affiliation{%
  \institution{Alibaba Group}
  \city{Beijing}
  \country{China}}
\email{liushuliang.lsl@alibaba-inc.com}

\author{Kaikui Liu}
\affiliation{%
  \institution{Alibaba Group}
  \city{Beijing}
  \country{China}}
\email{damon@alibaba-inc.com}

\author{Fan Yang}
\affiliation{%
  \institution{Alibaba Group}
  \city{Beijing}
  \country{China}}
\email{yangfan.yang@alibaba-inc.com}

\author{Xiangxiang Chu}
\affiliation{%
  \institution{Alibaba Group}
  \city{Beijing}
  \country{China}}
\email{xiangxiang.chucxxgtxy@gmail.com}

\author{Ning Guo}
\affiliation{%
  \institution{Alibaba Group}
  \city{Beijing}
  \country{China}}
\email{ning.guo@alibaba-inc.com}

\renewcommand{\shortauthors}{Jiahao et al.}

\begin{abstract}
  Multi-scenario route ranking (MSRR) is crucial in many industrial mapping systems. However, the industrial community mainly adopts interactive interfaces to encourage users to select pre-defined scenarios, which may hinder the downstream ranking performance. In addition, in the academic community, the multi-scenario ranking works only come from other fields, and there are no works specifically focusing on route data due to lacking a publicly available MSRR dataset. Moreover, all the existing multi-scenario works still fail to address the three specific challenges of MSRR simultaneously, \emph{i.e.} explosion of scenario number, high entanglement, and high-capacity demand. Different from the prior, to address MSRR, our key idea is to factorize the complicated scenario in route ranking into several disentangled factor scenario patterns. Accordingly, we propose a novel method, \textbf{D}isentangled \textbf{S}cenario \textbf{F}actorization \textbf{Net}work (\model{}), which flexibly composes scenario-dependent parameters based on a high-capacity multi-factor-scenario-branch structure. Then, a novel regularization is proposed to induce the disentanglement of factor scenarios. Furthermore, two extra novel techniques, \emph{i.e.} scenario-aware batch normalization and scenario-aware feature filtering, are developed to improve the network awareness of scenario representation. Additionally, to facilitate MSRR research in the academic community, we propose \dataset{}, the first large-scale publicly available annotated industrial \textbf{M}ulti-\textbf{S}cenario \textbf{D}riving \textbf{R}oute dataset. Comprehensive experimental results demonstrate the superiority of our \model{}, which has been successfully deployed in AMap to serve the major online traffic.
\end{abstract}

\ccsdesc[500]{Information systems~Recommender systems}
\ccsdesc[500]{Information systems~Retrieval models and ranking}

\keywords{Multi-Scenario Learning, Route Ranking, Recommender System, Route Recommendation, Scenario Disentanglement}



\maketitle

\section{Introduction}

Route recommendation is one of the core functionalities in many online location-based applications, which aims to predict the most likely route given an origin and a destination on the road network. Many industrial mapping systems (\emph{e.g.} AMap, Apple Maps, Baidu Maps, Tencent Maps) have started to build a two-stage recall\&rank framework to provide route recommendation service for users, where a lightweight efficient routing algorithm (\emph{e.g.} A* search) is applied to recall multiple \emph{feasible} routes in the first stage and a series of complex ranking models are deployed to further rank the recalled routes to screen out several \textit{user-preferred} ones in the second stage. This two-stage framework brings many benefits. For example, it ensures real-time recommendation services even if using deep learning techniques to model personalization. Moreover, it structurally decouples the modeling of feasibility and personalization of routes, which facilitates algorithm iteration and system troubleshooting. This paper focuses on the \textit{route ranking} stage.

When building the route ranking system, we found that the underlying \textit{multi-scenario} characteristics of user preference are critical for ranking user-preferred routes. Specifically, one may prefer different routes in different scenarios. For example, people who commute short distances in the morning rush hour prefer roads with shorter time and higher punctuality rates because they care much about time, while people who travel long distances during holidays prefer roads with more highways and fewer traffic lights because they focus much on driving comfort. This motivates us to study the problem of multi-scenario route ranking (MSRR).

Unfortunately, MSRR has not been well analyzed and addressed in both the academic and industrial communities. In the industrial community, mapping systems mainly define several scenarios manually before building the ranking model, and encourage users to select the scenario they are in by clicking the scenario click box on the apps page. However, the manual division is sub-optimal and may even hinder the downstream ranking tasks. Additionally, many users usually ignore click prompts, which brings noise to the scenario label in data collection for model iteration and may even impair the recommendation quality. In the academic community, there are even no works studying the MSRR problem, since the community lacks a publicly available dataset for MSRR. Fortunately, there are still many successful multi-scenario ranking methods~\cite{ple,star21,msr22/success,mb23/musenet}, though they come from other fields, where ranking objects may be products, ads, \emph{etc}. Nevertheless, all these works still fail to properly solve the MSRR problem. To explain the reason, we first exhaustively analyze the unique characteristics of the multi-scenario route data, which is divided into four points:
\begin{itemize}
    \item[1)] Scenario representation of the route is \textbf{fine-grained} and non-categorical (\emph{e.g.} departure time, status of road network between origin and destination). This is different from the explicitly categorical scenario in the existing multi-scenario ranking studies, which represents a pre-defined spot where products (or ads, videos, \emph{etc.}) are presented to users.
    \item[2)] Since fine-grained scenario representation values are mostly continuous in a wide range, it results in \textbf{explosion of scenario number} (\emph{abbr.} ESN) that combining different values of scenario representation to manually partition scenarios, which still cannot be avoided even with discretization.
    \item[3)] Since fine-grained scenario representation is rich in semantics, it leads to the phenomenon of \textbf{high entanglement}. Specifically, we observed that users might focus on identical factors under highly different scenarios. For example, users in night commuting and long-distance travel scenarios both focus on the quality of routes, while users in morning commuting and train catching scenarios both prefer the least time-consuming routes. Thus exploring \emph{disentangled factors of scenario that affect user preference} is definitely necessary for tackling the MSRR problem.
    \item[4)] Route features are highly abundant since a route is not only described from various aspects (\emph{e.g.} distance, traffic, toll) but also composed of a long road link sequence. It implies that the route ranking task requires a model with large model capacity, namely \textbf{high-capacity} demand. 
\end{itemize}

\begin{figure}[tbp]

  \centering
   \includegraphics[width=1\linewidth]{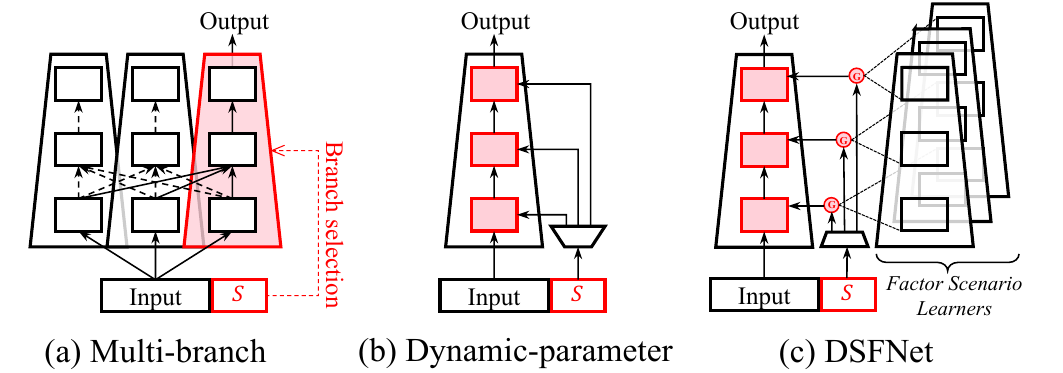}
   \caption{Comparison for the schemes of multi-scenario ranking method. (a) Maintain as many network branches as scenario number and one scenario-specific branch is selected in a forward propagation. (b) Directly generate the dynamic scenario-specific parameters of the network based on the scenario feature $S$. (c) Our method by composing dynamic parameters in a multi-factor-scenario-branch style.}
   \label{fig:compare}
\end{figure}

The above data characteristics introduce three specific challenges needing to be addressed in dealing with MSRR, \emph{i.e.} ESN problem, high entanglement, and high-capacity demand. However, all the existing multi-scenario ranking studies fail to address them simultaneously. Specifically, multi-scenario ranking solutions all design a model with scenario-dependent parameters, which can be mainly divided into two categories, \emph{i.e.} multi-branch method and dynamic-parameter method (\cref{fig:compare}). Despite multi-branch methods having high model capacity~\cite{moe}, the huge memory consumption caused by the ESN problem makes them impractical for industrial systems. In contrast, dynamic-parameter methods employ a much more flexible architecture to tackle the ESN problem. However, directly learning very high-dimensional dynamic parameters from scenario conditions is sub-effective in pattern learning and inefficient in time and memory ~\cite{APG/NEURIPS2022_9cd0c571}. Thus, Yan et al.~\shortcite{APG/NEURIPS2022_9cd0c571} recently utilized low-rank decomposition to improve these works. However, the low-rank constraint reduces the model capacity significantly, thus failing to meet the high-capacity demand. Moreover, all these works lack a module tailored to handle the highly-entangled scenario data.

Accordingly, for the first time, this paper proposes a novel methodology specifically for tackling the MSRR problem. Concretely, we first propose scenario factorization from a parametric perspective, \emph{i.e.} decomposing scenario-dependent parameters into several shared parameter sets, each of which represents a factor scenario learner (FSL). The scenario representation is responsible for generating suitable gates for all FSLs. This method essentially composes dynamic parameters in a multi-factor-scenario-branch structure, thus not only solving the ESN problem but also fulfilling the high-capacity demand. Then, to resolve high entanglement, we propose a novel disentangling regularization, which consists of two terms, \emph{i.e.} neuron centroid repulsion (NCR) and contrastive neuron clustering (CNC), to learn a semantically disentangled factor scenario for each FSL in a layer-wise manner. For each layer of paralleled FSLs, NCR keeps the neuron centroids as far away from each other as possible and CNC pushes directions of all neuron vectors in each FSL to cluster appropriately in parameter hyperspace, which not only induces the interpretability of factor scenarios (\cref{subsec:results}), but also improve the model generalization~\cite{MMA20/NEURIPS2020_dcd2f3f3}. In addition, we provide two extra novel techniques, scenario-aware batch normalization (SABN) and scenario-aware feature filtering (SAFF), which significantly improve the model performance since they both enhance the network awareness of the scenario feature. Overall, the four techniques proposed essentially inject the scenario information into all the locations of an MLP with learnable parameters. Specifically, scenario factorization and disentangling regularization enhance layer-wise weights and biases, SAFF boosts the feature embedding, and SABN strengthens layer-wise batch normalization.

Moreover, since the lack of a publicly available dataset is a major obstacle to MSRR research in the academic community, we propose \dataset{}, a large-scale publicly available annotated industrial multi-scenario driving route dataset. \dataset{} was collected when AMap, a top-tier location-based service provider in China, provided driving navigation services for users. In \dataset{}, each data sample includes route features, fine-grained scenario representation, user profiles, and a binary label representing whether the user prefers this route.

The major contributions can be summarized as follows.

\begin{itemize}
    \item We are the first to specifically analyze the multi-scenario route ranking problem (MSRR) brought from the industrial background and to propose a novel method, \textbf{D}isentangled \textbf{S}cenario \textbf{F}actorization \textbf{Net}work (\model{}), as a tailored solution to overcome the specific challenges of MSRR.
    
    \item We propose four novel techniques, \emph{i.e.} scenario factorization, disentangling regularization, scenario-aware batch normalization (SABN), and scenario-aware feature filtering (SAFF), to arm a vanilla MLP into our \model{}. And we theoretically analyze the advantages of the formulation of our disentangling regularization.
    
    \item We propose \dataset{}, the first large-scale publicly available annotated industrial \textbf{M}ulti-\textbf{S}cenario \textbf{D}riving \textbf{R}oute dataset, to facilitate MSRR research in the academic community\footnote{Data and codes will be released after this paper is published.}.
    
    \item We conduct extensive experiments and both offline and online experiments demonstrate the superiority of our \model{}. \model{} has been successfully deployed in AMap to serve the major online traffic.
    
\end{itemize}

\section{Related Work}

\subsection{Multi-Scenario Ranking}

Data distributions under different scenarios are quite different, which is hard for a common model to handle. The existing multi-scenario ranking methods all introduce scenario-dependent parameters to model data under different scenarios, which mainly fall into two categories, \emph{i.e.} multi-branch method and dynamic-parameter method. Multi-branch method designs multiple network branches based on the multi-task learning framework~\cite{shareBottom,mmoe,hmoe,ple}. Each branch is responsible for modeling data distribution of one specific scenario, and various sharing mechanisms are designed at the branch bottom to transfer inter-scenario knowledge~\cite{mb21/sar-net,mb22/sass,mb22/zou-2022-automatic}. Recently, STAR~\cite{star21} designed shared parameter gating to learn common knowledge. HiNet~\cite{zhou2023hinet} introduced a hierarchical branch and utilized an attentive module to model scenario correlation. Despite the high model capacity, the number of parameters of these works grows linearly with the scenario number, which is extremely inflexible. To tackle this, the dynamic-parameter method directly generates dynamic scenario-dependent parameters of the network from scenario features~\cite{dp22/m2m,dp22/CAN,dp22/causalint,dp23/digmn}. However, their parameter 
generation process lacks a shared component, which may cause model ignorance of the common patterns, leading to sub-effective pattern learning~\cite{APG/NEURIPS2022_9cd0c571}. Recently, Yan et al.~\shortcite{APG/NEURIPS2022_9cd0c571} exploited low-rank constraint to decompose parameters into dynamic low-rank kernels and shared weights to improve the effectiveness and efficiency of these works. Though effective, all these works are tailored for ranking objects other than routes, \emph{e.g.} products, ads, videos, and do not consider the unique characteristics of the multi-scenario route data, thus failing to properly solve the MSRR problem.

\subsection{DSFNet \emph{vs.} MoE \& MuSeNet} 
The series of classic MoE~\cite{moe,eigen2013learning/moe-series,shazeer2017outrageously/moe-series} and the recent MuSeNet~\cite{mb23/musenet} are both superficially similar to our \model{} since they all have multi-branch structures and gating mechanisms. But they have at least three differences: 
\begin{itemize}
    \item[1)] \emph{Different gating targets}: The gating targets of MoE and MuSeNet are both features (\emph{i.e.} outputs of branches), whereas our targets are network parameters (\emph{i.e.} FSL). Actually, since parameters are fixed (\emph{i.e.} input-independent), \model{} builds an interpretable scenario space where FSL represents coordinate basis and gating represents scenario point (\cref{subsubsec:interpret}), and that is what scenario factorization implies.
    \item[2)] \emph{Different motivations of multi-branch}: MoE expects each expert branch to only handle a subset of data to reduce the learning difficulty of each branch, thus accelerating the overall learning~\cite{moe}. MuSeNet aims to model specific user intention under each implicit \textbf{centroid} scenario. Our \model{} decomposes parameters into several branches to learn factor scenario patterns (\emph{i.e.} coordinate \textbf{bases}). 
    \item[3)] \emph{Different constraints on multi-branch}: MoE does not set constraints. MuSeNet uses prototypical clustering~\cite{pcl/li2021prototypical} to induce semantic implicit scenario prototype vectors, while our \model{} learns disentangled factor scenarios with a novel disentangling regularization on parameters.
\end{itemize}

\section{Method}

We formulate MSRR as a click-through-rate (CTR) prediction problem. Mathematically, given a set of routes $\mathcal{R}$ from the previous recalling stage, for a user $u \in \mathcal{U}$, the goal of MSRR is to infer the route $r \in \mathcal{R}$ most likely to be preferred by the user $u$ considering route features $f_r$, user profiles $f_u$, and scenario representation $f_s$, formally defined as solving the optimal route with the highest conditional probability:
\begin{equation} \label{eq:probForm}
    r^* = \mathop{\arg\max}_{r\in \mathcal{R}} \mathbf{P}(r | f_r, f_u, f_s)
\end{equation}

\subsection{Scenario Factorization Network}

\begin{figure*}[tbp]
  \centering
   \includegraphics[width=0.8\linewidth]{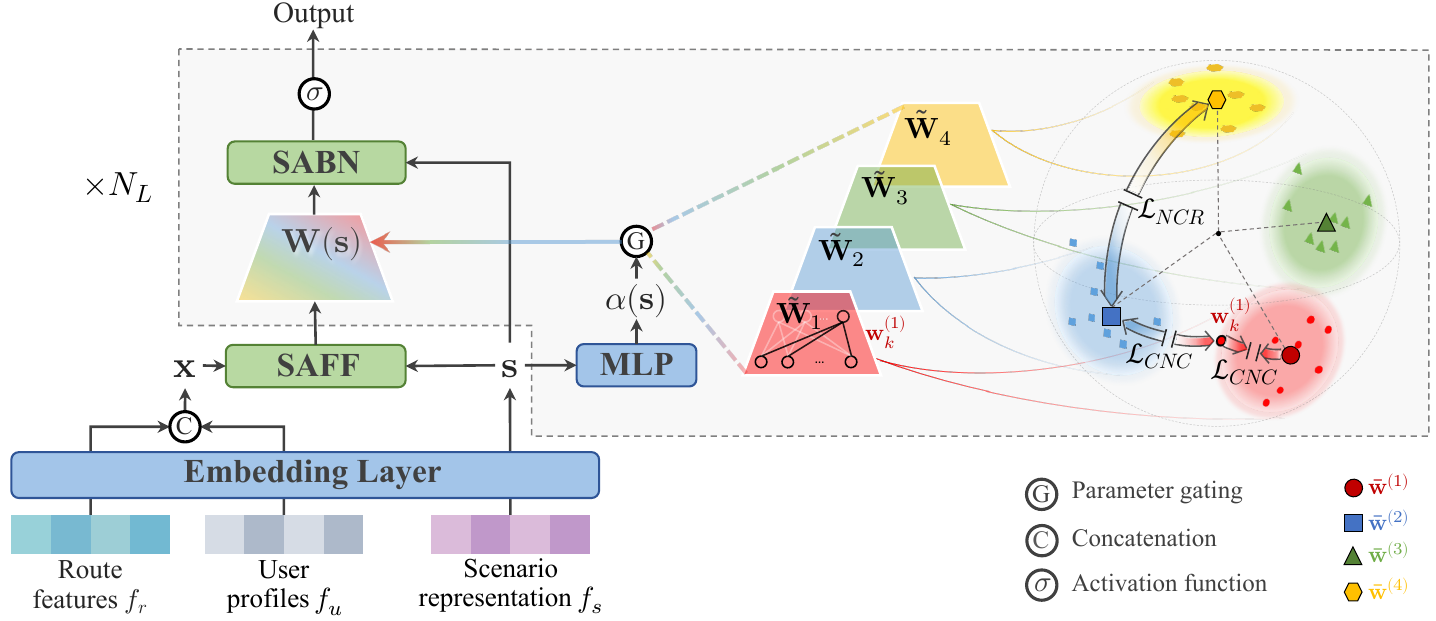}

   \caption{The illustration of \model{} with four FSLs (\emph{i.e.} $N=4$). In each layer, scenario factorization structure composes the dynamic network parameter $\mathbf{W}(\mathbf{s})$ via gating the four shared factor parameters $\{ \tilde{\mathbf{W}}_i \}_{i=1}^4$ from FSLs with the gates $\alpha(\mathbf{s})$. The disentangling regularization constrains the four factor parameters by $\mathcal{L}_{NCR}$ keeping the neuron centroids $\{ \bar{\mathbf{w}}^{(i)} \}_{i=1}^4$ as far away from each other as possible and $\mathcal{L}_{CNC}$ pushing the directions of the neurons in each $\tilde{\mathbf{W}}_i$ to cluster appropriately (\cref{subsec:DR}). SABN and SAFF are integrated to enhance the network awareness of $\mathbf{s}$ (\cref{subsec:SA}). Bias is omitted for figure simplicity.}
   \label{fig:arch}
\end{figure*}

To model the conditional probability in \cref{eq:probForm}, one common approach is to utilize a neural network $\mathbf{f}(\mathbf{x},\mathbf{s}; \mathbf{\Theta})$, where $\mathbf{x}$ is the concatenation of the embedding of $f_r$ and $f_u$, $\mathbf{s}$ is the embedding of $f_s$ and $\mathbf{\Theta}$ is the network parameter set. To better handle multi-scenario characteristics, 
multi-branch method redesigns the modeling as $\sum_{i=1}^{N_s} \mathbf{y_i}(\mathbf{s}) \mathbf{f}(\mathbf{x},\mathbf{s}; \mathbf{\Theta_i})$, where $N_s$ is scenario number and $\mathbf{y_i}(\mathbf{s})$ is the $i$ 'th element of the one-hot scenario indicating vector $\mathbf{y}(\mathbf{s})$. And dynamic-parameter method redesigns the modeling as $\mathbf{f}(\mathbf{x};\mathbf{\Theta}(\mathbf{s}))$, where $\mathbf{\Theta}(\cdot)$ is typically an MLP. Different from the prior work, we propose scenario factorization from a parametric perspective, \emph{i.e.} 
redesigning the modeling as $\mathbf{f}(\mathbf{x};\sum_{i=1}^{N} \alpha_{i}(\mathbf{s})\mathbf{\tilde{\Theta}_i})$, where $N$ is a hyperparameter, $\alpha(\cdot)$ is a gating network and $\mathbf{\tilde{\Theta}_i}$ is named as factor scenario learner (FSL). As shown in \cref{fig:arch}, the dynamic parameters in the $l$ 'th layer of our network are generated via
\begin{equation}
    \left[ \mathbf{W}^{(l)}(\mathbf{s}), \mathbf{b}^{(l)}(\mathbf{s})\right]=\sum_{i=1}^{N} \alpha^{(l)}_i(\mathbf{s})\left[\mathbf{\tilde{W}}_i^{(l)}, \mathbf{\tilde{b}}_i^{(l)}\right]
\end{equation}
\begin{equation}
    \alpha^{(l)}(\mathbf{s})=\sigma(\mathrm{MLP}^{(l)}(\mathbf{s}))
\end{equation}
where $\sigma(\cdot)$ is the sigmoid activation and the gating network $\alpha^{(l)}(\cdot)$ shares parameters across all layers in implementation.

\subsection{Disentangled Factorization}\label{subsec:DR}
\subsubsection{Overview.}
Scenario factorization without additional priors cannot spontaneously mine underlying disentangled factors from the highly entangled data. Even if it could, multiple FSLs would redundantly fit some common factor scenarios, \emph{e.g.} time-concerned scenario, because neural networks could be easily biased towards the head distribution of data \cite{longTail}. To address these issues, we first propose our Assumption 3.1 as follows:
\begin{assumption}\label{assumption}
    Keeping parameters of FSLs as angularly separate as possible can induce FSLs to mine disentangled factor scenarios.
\end{assumption}
This is inspired by the observation from the studies on angular diversity \cite{MHE/NEURIPS2018_177540c7,MMA20/NEURIPS2020_dcd2f3f3} that neural networks with angularly separated neurons in each layer are prone to capture different aspects of the data. Based on our assumption, we propose our disentangling regularization by composing two novel regularization terms, \emph{i.e.} neuron centroid repulsion (NCR) and contrastive neuron clustering (CNC), to achieve disentangled factorization in a layer-wise manner. Specifically, in the $l$ 'th layer of paralleled FSLs, NCR keeps the neuron centroids, each of which represents a $\mathbf{\tilde{W}}_i^{(l)}$, as far away from each other as possible, and CNC pushes the directions of the neurons in each $\mathbf{\tilde{W}}_i^{(l)}$ to cluster appropriately. The subsequent analysis is carried out for the $l$ 'th layer, thus we use $\mathbf{{W}}^{(i)}$ to denote $\mathbf{\tilde{W}}_i^{(l)}$ for simplicity. We define the neuron centroid as $
\bar{\mathbf{w}}^{(i)}=\sum_{k=1}^M \mathbf{w}_k^{(i)}/ {\Vert \sum_{k=1}^M \mathbf{w}_k^{(i)} \Vert} 
$, where $\mathbf{w}_k^{(i)} \in \mathbb{R}^{d\times 1}$ is the $k$ 'th neuron, \emph{i.e.} the transposed row vector, of $\mathbf{{W}}^{(i)}$, $M$ is the row number, \emph{i.e.} output dimension of this layer, and $\Vert \cdot \Vert$ denotes the Euclidean norm.

\subsubsection{Neuron centroid repulsion.} 
Keeping neuron centroids as far away from each other as possible is equivalent to maximizing the minimal pairwise distance between the centroids, which is known as the Tammes problem~\cite{tammes}. As the centroids are normalized, this problem can be recast into $\min \max_{i\neq j} {\bar{\mathbf{w}}^{(i)}}\cdot\bar{\mathbf{w}}^{(j)}$. Inspired by \cite{MMA20/NEURIPS2020_dcd2f3f3}, we first define an adversarial loss $\mathcal{L}_{MMA}$ to solve the minimax:
 \begin{equation}\label{eq:mma}
    \mathcal{L}_{MMA}=-\frac{1}{N}\sum_{i=1}^N \min_{j\neq i}\bm{\theta}^{(i,j)}
\end{equation}
where $\bm{\theta}^{(i,j)}$ is the angle between $\bar{\mathbf{w}}^{(i)}$ and $\bar{\mathbf{w}}^{(j)}$. However, we find that \cref{eq:mma} interferes with the convergence of the main CTR prediction task after the convergence of $\mathcal{L}_{MMA}$, which is not expected. Thus, we analyze the gradient of $\mathcal{L}_{MMA}$:
\begin{equation} 
    \begin{aligned}
    & \Vert \frac{\partial \mathcal{L}_{MMA}}{\partial \mathbf{w}_k^{(i)}} \Vert \propto \Vert \frac{\partial \bm{\theta}^{(i,j)}}{\partial \mathbf{w}_k^{(i)}} \Vert 
    = \Vert (\frac{\partial \bar{\mathbf{w}}^{(i)}}{\partial \mathbf{w}_k^{(i)}})^T \frac{\partial\bm{\theta}^{(i,j)}}{\partial \bar{\mathbf{w}}^{(i)}} \Vert \\
    & = \frac{1}{\Vert \sum_{l=1}^M \mathbf{w}_l^{(i)} \Vert}, \quad
     \mathrm{s.t.} \quad j=\argmin_{j\neq i,1\leq j\leq N} \bm{\theta}^{(i,j)}
\end{aligned}
\end{equation}
 Although the angular-independent gradient norm is beneficial for solving gradient vanishing when $\bm{\theta}^{(i,j)}$ is close to zero at the early stage of training~\cite{MMA20/NEURIPS2020_dcd2f3f3}, it becomes a gradient noise after the convergence of $\mathcal{L}_{MMA}$ that interferes with the model learning on the main CTR prediction task. To induce the gradient norm to converge to zero as neuron centroids move far away enough from each other, we propose neuron centroid repulsion (NCR) in \cref{eq:NCR}.
\begin{equation}\label{eq:NCR}
    \mathcal{L}_{NCR}=\frac{1}{N}\sum_{i=1}^{N}{\max_{j\neq i}{\Vert \bm{\theta}^{(i,j)}-\arccos(-\frac{1}{N-1}) \Vert^2}}
\end{equation}
which is derived from our Lemma 3.1, proven in \cref{app:proof}.
\begin{lemma} \label{lemma}
When $d\geq N-1$, $\min \max_{i\neq j} {\bar{\mathbf{w}}^{(i)}}\cdot\bar{\mathbf{w}}^{(j)} \Leftrightarrow 
\forall_{i\neq j}\,\bm{\theta}^{(i,j)} \\ = \arccos(-\frac{1}{N-1})
\Leftrightarrow
\min \max_{i\neq j} \Vert \bm{\theta}^{(i,j)}-\arccos(-\frac{1}{N-1})\Vert^2$.
\end{lemma}

The gradient norm of NCR is further analyzed in \cref{eq:NCR_grad}. It shows that not only NCR inherits the ability from $\mathcal{L}_{MMA}$ of solving the gradient vanishing, but also the gradient norm gradually converges to zero as NCR converges, which no longer introduces noises to interfere with the model learning of the main CTR prediction task.
\begin{equation}\label{eq:NCR_grad}
    \begin{aligned}
        \Vert \frac{\partial \mathcal{L}_{NCR}}{\partial \mathbf{w}^{(i)}_k} \Vert 
        & \propto   |\bm{\theta}^{(i,j)}-\arccos(-\frac{1}{N-1})|\times \Vert \frac{\partial \bm{\theta}^{(i,j)}}{\partial \mathbf{w}_k^{(i)}}\Vert \\
        & =\frac{\vert \bm{\theta}^{(i,j)}-\arccos(-\frac{1}{N-1})\vert}{\Vert \sum_{l=1}^M \mathbf{w}_l^{(i)} \Vert}, \\
        \mathrm{s.t.}\quad j= & \argmax_{j\neq i,1\leq j\leq N}{\Vert \bm{\theta}^{(i,j)}-\arccos(-\frac{1}{N-1}) \Vert^2}
    \end{aligned}
\end{equation}
\subsubsection{Contrastive neuron clustering.}

Pushing the directions of neurons in $\mathbf{W}^{(i)}$ to cluster appropriately is conducive to learning disentangled factors, whereas pushing them too close to each other will lead to poor generalization due to the lack of angular diversity. To prevent the directions of neurons from being too close, we propose contrastive neuron clustering (CNC) in \cref{eq:CNC} with the idea of contrastive learning, where $\bm{\theta}\langle a,b\rangle$ denotes the angle between $a$ and $b$, and $\delta_\theta$ is a hyperparameter. CNC first denotes the neuron $\mathbf{w}^{(i)}_k$ farthest from $\bar{\mathbf{w}}^{(i)}$ (\emph{i.e.} positive) in $\mathbf{W}^{(i)}$ as an anchor, then constrains the angle between the anchor and the positive smaller than the angle between the anchor and the closest $\bar{\mathbf{w}}^{(j)}$ (\emph{i.e.} negative) to some extend (measured by $\delta_\theta$). 
\begin{equation} \label{eq:CNC}
    \begin{aligned}
        \mathcal{L}_{CNC}=\frac{1}{N}\sum_{i=1}^N &
    \max(0, \delta_\theta +
    \bm{\theta}\langle\mathbf{w}^{(i)}_k,\bar{\mathbf{w}}^{(i)}\rangle-
    \bm{\theta}\langle\mathbf{w}^{(i)}_k,\bar{\mathbf{w}}^{(j)}\rangle),
    \\
    \mathrm{s.t.}\quad & k=\argmax_{1\leq k \leq M} \bm{\theta}\langle\mathbf{w}^{(i)}_k,\bar{\mathbf{w}}^{(i)}\rangle, \\
    & j =  \argmin_{j\neq i, 1\leq j\leq N} \bm{\theta}\langle\mathbf{w}^{(i)}_k,\bar{\mathbf{w}}^{(j)}\rangle
    \end{aligned}
\end{equation}

In practice, it is more efficient to replace angle with cosine similarity since the latter can be efficiently implemented with the dot product. However, using cosine similarity in CNC leads to the vanishing of the repulsion from negative to anchor, \emph{i.e.} gradient vanishing, especially in the early stage of training. Specifically, we analyze the gradient:
\begin{equation}\label{eq:CNC_cos_grad}
    \Big\Vert \frac{\partial \cos \bar{\bm{\theta}}_k^{(i,j)}}{\partial \mathbf{w}^{(i)}_k}\Big\Vert
    =\Big\Vert \frac{\partial\big( 
        \frac{\mathbf{w}^{(i)}_k \cdot \bar{\mathbf{w}}^{(j)}}
        {\Vert \mathbf{w}^{(i)}_k\Vert}
    \big)
    }{\partial \mathbf{w}^{(i)}_k} \Big\Vert
    =\frac{\sin \bar{\bm{\theta}}_k^{(i,j)}}{\Vert \mathbf{w}^{(i)}_k \Vert}
\end{equation}
where $\bar{\bm{\theta}}_k^{(i,j)}$ denotes the angle between $\mathbf{w}^{(i)}_k$ and $\bar{\mathbf{w}}^{(j)}$. It shows that the gradient norm is close to zero when $\bar{\bm{\theta}}_k^{(i,j)}$ (\emph{i.e.} the angle between negative and anchor) is small, which is prevalent in neural network\cite{prevalent}, especially when $\bar{\mathbf{w}}^{(j)}$ is the closest negative of the anchor $\mathbf{w}^{(i)}_k$. In contrast, CNC in angle form provides a gradient norm independent of $\bar{\bm{\theta}}_k^{(i,j)}$ (\cref{eq:CNC_grad}).
\begin{equation}\label{eq:CNC_grad}
    \Vert \frac{\partial \bar{\bm{\theta}}_k^{(i,j)}}{\partial \mathbf{w}^{(i)}_k}\Vert
    =\Vert \frac{\partial\bar{\bm{\theta}}_k^{(i,j)}}{\partial\cos \bar{\bm{\theta}}_k^{(i,j)}}
    \frac{\partial\cos \bar{\bm{\theta}}_k^{(i,j)}}{\partial \mathbf{w}^{(i)}_k}\Vert
    =\frac{1}{\Vert \mathbf{w}^{(i)}_k\Vert}
\end{equation}

\subsubsection{Disentangling regularization.}
In practice, NCR and CNC can be efficiently implemented with matrix multiplication. Accordingly, we develop the disentangling regularization in \cref{eq:reg} for training \model{}, where $N_L$ denotes the layer number of our \model{}. Thus, the total loss is shown in \cref{eq:tot_loss}, where $\mathcal{L}_{BCE}$ is binary cross entropy loss for CTR modeling and $\lambda$ is the regularization coefficient.
\begin{equation}\label{eq:reg}
    \mathcal{L}_{DR}=\sum_{l=1}^{N_L}
    {\mathcal{L}_{NCR}(\{ \mathbf{\tilde{W}}_i^{(l)} \}^N_{i=1})+
     \mathcal{L}_{CNC}(\{ \mathbf{\tilde{W}}_i^{(l)} \}^N_{i=1})
     }
\end{equation}
\begin{equation}\label{eq:tot_loss}
    \mathcal{L}=\mathcal{L}_{BCE}+\lambda\mathcal{L}_{DR}
\end{equation}

\subsection{Enhancing Awareness of Scenario} \label{subsec:SA}

Additionally, as the dimension of the route feature is much higher than that of the scenario feature, the network is prone to capture unreliable intra-correlation of the route feature, leading to weakened network awareness of the scenario feature~\cite{whytrust/unreliable}. To enhance the awareness, we further propose two practical techniques, \emph{i.e.} scenario-aware batch normalization and scenario-aware feature filtering, to inject more feature interactions to \model{}.
\subsubsection{Scenario-aware batch normalization.}
Batch normalization applies a global normalization and learnable re-scaling across all samples~\cite{BN/batchnorm}, which may be harmful to multi-scenario data as data distributions under different scenarios are different. \cite{star21} designed partitioned normalization (PN) to normalize and re-scale samples privately within each scenario. Despite the effectiveness of addressing the differences in scenarios, it requires that samples in a mini-batch are sampled from the same scenario during training, which is inflexible. And it learns a set of parameters for each scenario, which is impractical due to the ESN problem. Therefore, we propose scenario-aware batch normalization (SABN):
\begin{equation}
    z^{'} = \bm{\gamma}(\mathbf{s}_z)\frac{z-\mu_z}{\sqrt{\sigma_z^2+\epsilon}}+\bm{\beta}(\mathbf{s}_z)
\end{equation}
where $\mu_z$ and $\sigma_z^2$ are mean and variance of $z$ in current batch, $\bm{\gamma}(\cdot)$ and $\bm{\beta}(\cdot)$ are two MLPs, and $\mathbf{s}_z$ is scenario feature of sample $z$. Though the samples in a mini-batch are normalized by scenario-agnostic statistics, they are re-scaled by scenario-dependent parameters. Thus, it not only overcomes the limitations of PN but also induces the model to learn the feature distribution transformation from scenario-agnostic distribution to scenario-specific one. During testing, $\mu_z$ and $\sigma_z^2$ are respectively replaced with the moving average of mean and variance accumulated in the training phase.
\subsubsection{Scenario-aware feature filtering.} Furthermore, we seek to model the process of humans manually selecting features based on scenario information, which is common in the feature engineering phase. Inspired by the gating technique~\cite{gatenet}, we propose scenario-aware feature filtering (SAFF) via feeding $\mathbf{s}$ into the gating network:
\begin{equation}
    \mathbf{x}^{'}= \mathbf{x}\odot\sigma(\mathrm{MLP(\mathbf{x},\mathbf{s})})
\end{equation}
where $\sigma$ is sigmoid activation and $\odot$ denotes Hadamard product.

\section{The Proposed Dataset}
Since there are no available benchmarks for evaluating MSRR models in the academic community, we propose \textbf{M}ulti-\textbf{S}cenario \textbf{D}riving \textbf{R}oute dataset (\dataset{}), the first large-scale publicly available annotated industrial dataset, to facilitate MSRR research in the academia.
\subsection{Data Acquisition}\label{subsec:data_acq}
\dataset{} was collected when AMap, a top-tier location-based service provider in China, provided driving navigation services for users. Given the origin and destination, one navigation service may involve multiple route recommendation services, where the recommendation service is supported by a recall\&rank framework, and in each route recommendation service it retrieves up to one hundred candidate routes in the recalling stage and screens out three routes as the system outputs from the candidates in the ranking stage. 

Specifically, when a user determines the destination through AMap, the navigation service enters the standby state and initiates a route recommendation, and then the three output routes are exposed to the user through the app page. When the user selects one route, the navigation service is officially started. During the navigation, once the user deviates from the selected route, a new route recommendation will be triggered at the off-route position. AMap collects the selected route set, the off-route position set, the times of deviation in a navigation(ToD), the recalled routes in the first recommendation, and the corresponding scenario information. The navigation service lasts until the user reaches the destination or ends the navigation prematurely. To obtain the user's preference labels for routes, we use the selected route set and the off-track position set to splice the groundtruth route, which is used for computing the coverage rate with all the recalled routes. Then the route with the highest coverage rate among the recalled ones is labeled as positive, while the rest are all labeled as negative. AMap further collects the route labels and the coverage rate of each route into the database. To construct a label-balanced dataset, we collect only three or four negative routes in each navigation, \emph{i.e.} the non-positive exposed routes (two or three), and one randomly sampled from the other recalled routes. Moreover, navigation data that end prematurely are discarded to avoid introducing noise.

\subsection{\dataset{} Dataset}\label{subsec:dataset}
\begin{figure}[tbp]

  \centering
   \includegraphics[width=1\linewidth]{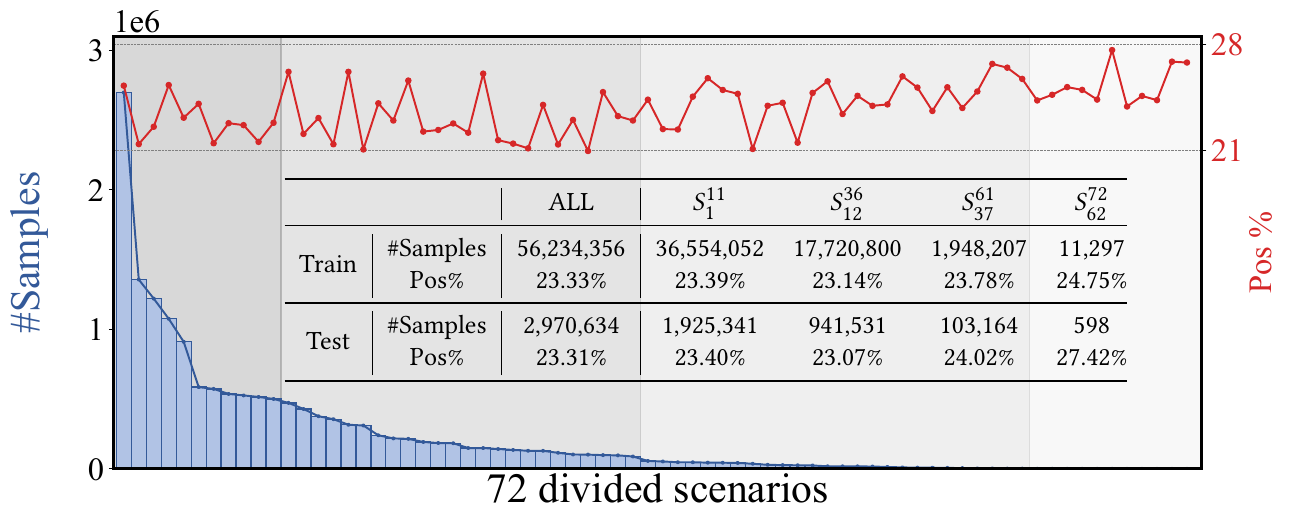}
   \caption{The number of samples (\#Samples) and proportion of the positives (Pos\%) of the 72 scenarios, which are further divided into four subsets (by four background colors), and the statistics of the subsets are exhibited in the table.}
   \label{fig:data}
\end{figure}

Through extracting samples from the data acquisition process, we obtained our \dataset{}. It contains two weeks of data, from June 25th, 2023 to July 8th, 2023, during which there are no Chinese holidays. We sampled data from eight big cities in China, \emph{i.e.} Beijing, Shanghai, Guangzhou, Hangzhou, Wuhan, Zhengzhou, Chongqing, and Chengdu. Each data sample includes abundant route features, fine-grained scenario representation, user profiles, and a preference label. Specifically, \cref{tab:routefeat} exhibits some route features provided in \dataset{}. The user profiles include gender, age, and occupation, which are all encrypted through hash and do not disclose privacy. The scenario representation includes time, date, some road network characteristics between origin and destination (\emph{i.e.} congestion situation), spherical distance between origin and destination, POI (Point-of-interest) category of destination, location type of origin and destination, and user familiarity with origin and destination. To adapt \dataset{} to the multi-branch methods, we select some scenario feature fields (\cref{tab:div}) to manually divide the scenario representation into $3 \times 2 \times 3 \times 2 \times 2 = 72$ scenarios. As shown in \cref{fig:data}, we further divide the long-tail data under the 72 scenarios into four subsets, \emph{i.e.} $S_1^{11}$,$S_{12}^{36}$, $S_{37}^{61}$, $S_{62}^{72}$, following the order from the high sample number to the low in a partition ratio of 15\%, 50\%, and 85\%, where $S_i^{j}$ denotes samples under a scenario subset including from the $i$-th scenario to the $j$-th scenario. This division allows us to study model performances on the head and tail scenarios, respectively. \dataset{} contains 59,204,990 samples and 10,461,969 users, and the proportion of positives in \dataset{} is $23.33\%$. We divide \dataset{} into a train set and a test set, whose statistics are shown in \cref{fig:data}. The `Pos\%' of the four subsets in \cref{fig:data} reveal that the head scenarios have higher inducing effect on users' deviation compared with the tail ones, \emph{e.g.} the head ones include the long-distance travel scenario which has a large quantity of data. Finally, we compare our dataset with some representative route datasets in \cref{tab:dataset}, including T-Drive~\cite{zheng2011t-drive/TD}, NYCT~\cite{NYC_taxi/NT}, PorT~\cite{Porto/pt/Starace2020} and LonT~\cite{LT/london_trajectory}.
\begin{table}[tb]
  \centering
  \scalebox{0.65}{
  \begin{tabular}{@{}l|l@{}}
    \toprule
    Route Feature & Description \\
    \midrule
    ETASecond & The estimated time of arrival for the route in seconds \\
    Distance & The total distance length of the route \\
    HighwayLen & The total length of highway roads in the route \\
    NonHWLen & The total length of non-highway roads in the route \\
    ArrHWDis & The distance to the highway from the beginning of the route \\
    Toll & The total toll amount for the route in Chinese fen \\
    FamlrRate & The familiarity ratio of all the road links on the route \\
    HeadFamlr & The sum of all the familiar links’ frequency at the beginning of the route \\
    \#TrafLights & The number of traffic lights in the route \\
    ArrUnkDis & The distance to roads with unknown conditions from the beginning of the route \\
    \#Forwards & The number of times to go straight in the route \\
    LeavUnkDis & The distance from roads with unknown conditions to the end of the route \\
    LeavHWDis & The distance from the highway to the end of the route \\
    AlleyLen & The total distance of roads with an extremely poor capacity on the route \\
    AvgKMPH & The average speed for the route in kilometers per hour \\
    \bottomrule
  \end{tabular}
  }
  \caption{Some examples of the route features provided in \dataset{}, along with the corresponding descriptions. }
  \label{tab:routefeat}
\end{table}

\begin{table}[tb]
  \centering
  \scalebox{0.65}{
  \begin{tabular}{@{}m{200pt}|m{140pt}@{}}
    \toprule
    Selected Scenario Feature Field & Manual Division\\
    \midrule
    Time & Morning rush, evening rush, or others \\
    Date & Weekdays or weekends \\
    Spherical distance between origin and destination & Long, intermediate, or short distance \\
    Congestion situation between origin and destination & Congested or not congested \\
    Proportion of highways between origin and destination  & Highway exists or not \\
    \bottomrule
  \end{tabular}
  }
  \caption{Empirical rules for manually dividing the scenarios. }
  \label{tab:div}
\end{table}

\begin{table}[htb]
  \centering
  \scalebox{0.8}{
  \begin{tabular}{@{}l|c|cccc@{}}
    \toprule
          & \textbf{\dataset{}} & T-Drive~\cite{zheng2011t-drive/TD} & PorT~\cite{Porto/pt/Starace2020} & NYCT~\cite{NYC_taxi/NT} & LonT~\cite{LT/london_trajectory} \\
    \midrule
    \#Users & \textbf{10.46M} & 33K & 442 & 39K & - \\
    \#Routes & 59.2M & 4.96M & 284K & \textbf{697.62M} & 600 \\
    \#Cities & \textbf{8} & 1 & 1 & 1 & 1 \\
    \midrule
    GPS & N & \textbf{\textcolor[RGB]{34,178,34}{Y}} & \textbf{\textcolor[RGB]{34,178,34}{Y}} & \textbf{\textcolor[RGB]{34,178,34}{Y}} & \textbf{\textcolor[RGB]{34,178,34}{Y}} \\
    RouF & \textbf{\textcolor[RGB]{34,178,34}{Y}} & N & N & N & \textbf{\textcolor[RGB]{34,178,34}{Y}} \\
    FinSc & \textbf{\textcolor[RGB]{34,178,34}{Y}} & N & N & N & N \\
    UsPr & \textbf{\textcolor[RGB]{34,178,34}{Y}} & N & N & N & N \\
    Rank & \textbf{\textcolor[RGB]{34,178,34}{Y}} & N & N & N & N \\
    \bottomrule
  \end{tabular}
  }
  \caption{Comparison of some properties between \dataset{} and the other datasets. \dataset{} provides abundant route features (RouF), fine-grained scenario representation (FinSc), user profiles (UsPr), and ranking labels (Rank), whereas \dataset{} does not provide any GPS information due to privacy concerns.}
  \label{tab:dataset}
\end{table}

\section{Experiments} \label{sec:exps}
\subsection{Experiment Settings}
\subsubsection{Datasets and metrics.} 
Since our \dataset{} is the only benchmark available for evaluating MSRR models, we conduct experiments on the \dataset{} dataset. For metrics, we apply AUC (Area Under the ROC) and RelaImpr \cite{relaimp0,relaimp1} for evaluation. We further calculate subset AUC, \emph{i.e.} AUC$_1^{11}$, AUC$_{12}^{36}$, AUC$_{37}^{61}$, and AUC$_{62}^{72}$, where AUC$_{i}^{j}$ denotes the AUC score calculated on $S_i^j$. In addition, we calculate the number of parameters of each model to compare memory complexity.

\begin{table*}[ht]
\centering
\scalebox{0.9}{
\begin{tabular}{@{}l|c|cc|cc|cc|cc|c|c@{}}
\toprule
Method                & \#Params & AUC(\%)        & RelaImpr        & AUC$_{1}^{11}$(\%) & RelaImpr        & AUC$_{12}^{36}$(\%) & RelaImpr        & AUC$_{37}^{61}$(\%) & RelaImpr        & AUC$_{62}^{72}$(\%) & RelaImpr        \\ \midrule
Recall                & -        & 50.16          & -32.08\%        & 50.36                          & -31.75\%        & 48.99                           & -33.77\%        & 52.17                           & -29.41\%        & \cellcolor[HTML]{EFEFEF}55.77                           & -22.80\%        \\
Vanilla MLP           & 0.11M    & 73.62          & 0.00\%          & 73.55                          & 0.00\%          & \cellcolor[HTML]{EFEFEF}73.71                          & 0.00\%          & 73.70                           & 0.00\%          & 72.09                           & 0.00\%          \\ \midrule
ShareBottom~\cite{shareBottom}          & 0.86M    & 74.08          & 0.63\%          & \cellcolor[HTML]{EFEFEF}74.17                          & 0.85\%          & 73.79                           & 0.11\%          & 73.66                           & -0.05\%         & 68.10                           & -5.57\%         \\
MMOE~\cite{mmoe}                  & 1.59M    & 74.15          & 0.72\%          & \cellcolor[HTML]{EFEFEF}74.18                          & 0.86\%          & 73.96                           & 0.34\%          & 73.65                           & -0.07\%         & 68.17                           & -5.48\%         \\
PLE~\cite{ple}                   & 15.44M   & 74.16          & 0.74\%          & \cellcolor[HTML]{EFEFEF}74.21                          & 0.90\%          & 73.93                           & 0.30\%          & 73.87                           & 0.23\%          & 68.43                           & -5.11\%         \\
STAR~\cite{star21}                  & 7.93M    & 74.13          & 0.70\%          & \cellcolor[HTML]{EFEFEF}74.18                          & 0.86\%          & 74.01                           & 0.41\%          & 73.72                           & 0.03\%          & 69.54                           & -3.56\%         \\
M2M~\cite{dp22/m2m}                   & 9.31M    & 73.67          & 0.07\%          & 73.57                          & 0.03\%          & 74.04                           & 0.45\%          & \cellcolor[HTML]{EFEFEF}74.17                           & 0.64\%          & 70.64                           & -2.03\%         \\
APG~\cite{APG/NEURIPS2022_9cd0c571}                  & 0.17M    & 72.91          & -0.97\%         & 72.89                          & -0.90\%         & 72.58                           & -1.54\%         & \cellcolor[HTML]{EFEFEF}73.37                           & -0.45\%         & 70.92                           & -1.63\%         \\
HiNet~\cite{zhou2023hinet}                 & 16.13M   & 74.41          & 1.08\%          & 74.35                          & 1.10\%          & \cellcolor[HTML]{EFEFEF}{\ul 74.65}                     & 1.28\%          & 74.46                           & 1.04\%          & 70.41                           & -2.35\%         \\
MuSeNet~\cite{mb23/musenet}               & 0.87M    & {\ul 74.52}    & 1.23\%          & {\ul 74.56}                    & 1.38\%          & 74.23                           & 0.71\%          & \cellcolor[HTML]{EFEFEF}{\ul 75.08}                     & 1.89\%          & {\ul 71.40}                     & -0.96\%         \\ \midrule
\textbf{DSFNet (Ours)} & 0.83M    & \textbf{75.04$^*$} & \textbf{1.94\%} & \textbf{74.96$^*$}                 & \textbf{1.93\%} & \textbf{75.72$^*$}                  & \textbf{2.75\%} & \cellcolor[HTML]{EFEFEF}\textbf{75.86$^*$}                  & \textbf{2.95\%} & \textbf{72.20$^*$}                  & \textbf{0.15\%} \\ \bottomrule
\end{tabular}
}
\caption{Comparison results of different methods on \dataset{} test. RelaImpr is calculated based on vanilla MLP. The best results are in boldface and the second best are underlined. \colorbox[RGB]{239,239,239}{Gray background color} indicates the result with the highest subset AUC among the four. $^*$ indicates that the difference to the best baseline is statistically significant at 0.05 level. }
  \label{tab:compare}
\end{table*}

\subsubsection{Competitors.} We first benchmark our \model{} against two baselines, which are not designed by the multi-scenario priors:
\begin{itemize}
    \item \textbf{Recall} represents the exclusive routing algorithm of the recalling stage in AMap's route recommendation service.
    \item \textbf{Vanilla MLP} is a naive multi-layer perceptron.
\end{itemize}
Then we compare our methods with some \textit{multi-branch} methods:
\begin{itemize}
    \item \textbf{ShareBottom}~\cite{shareBottom} builds scenario-specific tower networks on a shared bottom MLP to learn common knowledge.
    \item \textbf{MMOE}~\cite{mmoe} applies scenario-specific gates to the outputs of the shared experts to capture correlations among scenarios.
    \item \textbf{PLE}~\cite{ple} builds a shared branch and several scenario-specific ones, and applies a progressive routing mechanism to further decouple common knowledge and scenario-specific one.
    \item \textbf{STAR}~\cite{star21} applies a shared parameter gating to the parameters of scenario-specific branches to transfer common knowledge to the branches for scenario-specific domains.
    \item \textbf{HiNet}~\cite{zhou2023hinet} recently built scenario-specific branches and a shared one with a hierarchical structure and applied an attentive module to share knowledge across the scenarios.
\end{itemize}
Afterward, we compare our model with two \textit{dynamic-parameter} methods:
\begin{itemize}
    \item \textbf{M2M}~\cite{dp22/m2m} recently built a network based on meta units, which adopted an MLP architecture whereas its parameters were dynamically generated by the scenario attributes to learn explicit inter-scenario correlations.
    \item \textbf{APG}~\cite{APG/NEURIPS2022_9cd0c571} recently applied the low-rank constraint to decompose dynamic parameters into a small set of dynamic scenario-dependent parameters and a set of shared parameters to improve efficiency and effectiveness.
\end{itemize}
Finally, we compare our model with the recent MuSeNet~\cite{mb23/musenet}, which is also the most related work with ours:
\begin{itemize}
    \item \textbf{MuSeNet}~\cite{mb23/musenet} recently built a few branches to model data under the implicit scenario centroids, and exploited an alternating optimization to simultaneously perform scenario learning and ranking modeling.
\end{itemize}

\subsubsection{Implementation details.} We implement all methods using the Tensorflow toolbox on a Tesla V100 GPU with 16GB. Before fed into the model, $\mathbf{x}$ and $\mathbf{s}$ are normalized as $\hat{\mathbf{x}}$ and $\hat{\mathbf{s}}$ by moving mean and variance of batch data, respectively, except for the embedded categorical features. Besides $\mathbf{x}$ and $\mathbf{s}$, we have a one-hot scenario indicator $\mathbf{y}(\mathbf{s})\in \mathbb{R}^{72}$ from the scenario division. We concatenate $\mathbf{x}$ and $\mathbf{s}$ as the input for all multi-branch methods and MuSeNet, and $\mathbf{y}(\mathbf{s})$ is only used for branch selection in multi-branch methods, except that HiNet further exploits the embedding of $\mathbf{y}(\mathbf{s})$ according to \cite{zhou2023hinet}. In contrast, we only use $\mathbf{x}$  as the input for M2M, APG, and our \model{}, while $\mathbf{s}$ is only used for generating dynamic parameters and as input for SABN and SAFF. We train all models using the Adam optimizer~\cite{kingma2014adam} for 2M iterations on \dataset{} train. We set the batch size to 512. The learning rate is initially set to $0.001$ and it exponentially decays with a decay rate of $0.98$ and a decay step of 10K. In our \model{}, we use a five-layer MLP (\emph{i.e.} $N_L=5$) to model an FSL, whose hidden dimensions are 256, 128, 64, and 32, respectively. We set $N=7$, $\lambda=0.01$, and $\delta_\theta=\frac{1}{\kappa}\arccos(\frac{1}{1-N})$ with $\kappa=1.75$. In baselines, for fair comparison, we keep the depths and hidden dimensions of baselines identical to our \model{}. Specifically, we set the number of implicit scenarios to $N=7$ in MuSeNet. We thoroughly finetuned the low rank $K_{apg}$ in APG and finally set $K_{apg}=24$. We replace the multi-task layer in HiNet and M2M with an MLP with the same depth, respectively, since our problem is single-task. We keep the hidden dimension of M2M as 256 according to \cite{dp22/m2m}, thus resulting in the relatively high memory complexity.

\subsection{Results} \label{subsec:results}

\begin{table*}[ht]
\scalebox{0.95}{
\begin{tabular}{@{}cccc|c|cccc@{}}
\toprule
SF & SAFF & SABN & Regularization             & AUC(\%)        & AUC$_1^{11}$(\%) & AUC$_{12}^{36}$(\%)       & AUC$_{37}^{61}$(\%)       & AUC$_{62}^{72}$(\%)       \\ \midrule
\no  & \yes    & \yes    & -                          & 74.29 (\textcolor[RGB]{178,34,34}{-0.75})  & 74.24 (\textcolor[RGB]{178,34,34}{-0.72})                  & 74.54 (\textcolor[RGB]{178,34,34}{-1.18})                         & \cellcolor[HTML]{EFEFEF}74.51 (\textcolor[RGB]{178,34,34}{-1.35}) & 71.21 (\textcolor[RGB]{178,34,34}{-0.99})                         \\
\yes  & \no   & \yes    & $\mathcal{L}_{DR}$                         & 74.54 (-0.50)  & 74.52 (-0.44)                  & \cellcolor[HTML]{EFEFEF}74.93 (\textcolor[RGB]{178,34,34}{-0.79}) & \cellcolor[HTML]{EFEFEF}75.07 (\textcolor[RGB]{178,34,34}{-0.79}) & 71.46 (\textcolor[RGB]{178,34,34}{-0.74})                         \\
\yes  & \yes    & \no    & $\mathcal{L}_{DR}$                         & 74.86 (-0.18)  & 74.74 (-0.22)                  & 75.57 (-0.15)                         & \cellcolor[HTML]{EFEFEF}75.57 (-0.29) & 71.94 (-0.26)                         \\
\yes  & \yes    & \yes    & \no                          & 74.84 (-0.20)  & 74.91 (-0.05)                  & 74.93 (\textcolor[RGB]{178,34,34}{-0.79})                         & 74.75 (\textcolor[RGB]{178,34,34}{-1.11})                         & \cellcolor[HTML]{EFEFEF}70.91 (\textcolor[RGB]{178,34,34}{-1.29}) \\ \midrule
\yes  & \yes    & \yes    & $\mathcal{L}_{orth}$                       & 74.93 (-0.11)  & 74.90 (-0.06)                  & 75.67 (-0.05)                         & 74.94 (\textcolor[RGB]{178,34,34}{-0.92})                         & \cellcolor[HTML]{EFEFEF}70.85 (\textcolor[RGB]{178,34,34}{-1.35}) \\
\yes  & \yes    & \yes    & $\mathcal{L}_{MMA}+\mathcal{L}_{CNC}$                    & 74.82 (-0.22)  & 74.78 (-0.18)                  & 75.39 (-0.33)                         & 75.16 (\textcolor[RGB]{178,34,34}{-0.70})                         & \cellcolor[HTML]{EFEFEF}71.12 (\textcolor[RGB]{178,34,34}{-1.08}) \\
\yes  & \yes    & \yes    & $\mathcal{L}_{NCR}$                        & 74.92 (-0.12)  & 74.87 (-0.09)                  & 75.56 (-0.16)                         & \cellcolor[HTML]{EFEFEF}75.49 (-0.37) & \textbf{72.27} (+0.07)                         \\
\yes  & \yes    & \yes    & $\mathcal{L}_{NCR}+\mathcal{L}_{CNC}^{cos}$ & 74.96 (-0.08)  & 74.94 (-0.02)                  & 75.53 (-0.19)                         & \cellcolor[HTML]{EFEFEF}75.66 (-0.20) & 72.21 (+0.01)                         \\ \midrule
\yes  & \yes    & \yes    & $\mathcal{L}_{DR}$                         & \textbf{75.04} & \textbf{74.96}                 & \textbf{75.72}                        & \textbf{75.86}                        & 72.20                                 \\ \bottomrule
\end{tabular}
}
\caption{Ablation analysis of our four techniques, \emph{i.e.} scenario factorization (denoted as `SF'), disentangling regularization $\mathcal{L}_{DR}$, SABN, and SAFF, and comparison results of different regularizations. \no{} of `SABN' denotes replacing SABN with standard BN. The best results are in boldface. The number in the bracket denotes the gap with the full model, and in red are the drop gaps of at least \textcolor[RGB]{178,34,34}{-0.70\%}. \colorbox[RGB]{239,239,239}{Gray background color} indicates the result with the highest drop gap in subset AUC among the four.}
\label{tab:ablation}
\end{table*}

\subsubsection{Performance Comparison.} 
We report the comparison results on \dataset{} test in \cref{tab:compare}. As shown, our \model{} achieves the best performance on the overall \dataset{} and the four subsets, respectively, and it is the only model that achieves a positive gain over the MLP in the tail $S_{62}^{72}$. Moreover, it achieves the best trade-off between model complexity and performance over the previous methods. 

Then we analyze the baselines. For the recall baseline, the gap of about 20\% in AUC validates the indispensability of the ranking model, and the reason for the numerical relationship of the four subset AUCs can refer to the analysis for subset `Pos\%' in \cref{subsec:dataset}. For MLP, its performance is average but there is not much gap among the subset AUCs. For multi-branch methods, ShareBottom and MMOE both obtain improvements over the MLP, especially in the head $S_1^{11}$, $S_{12}^{36}$, whereas the seesaw phenomenon occurs as the performances worsen in the tail $S_{37}^{61}$, $S_{62}^{72}$. PLE and STAR alleviate the performance degradation in the tail $S_{37}^{61}$. HiNet further alleviates the issues and brings about AUC$_{12}^{36}$ reaching the highest among the subset AUCs, which is the best multi-branch method. However, some multi-branch methods (\emph{i.e.} PLE, HiNet) have nearly 20 times as many parameters as ours due
to the ESN problem. For dynamic-parameter methods, M2M exhibits average performance in the head $S_1^{11}$ probably because its dynamic parameter generation process is merely dependent on the scenario conditions, which may cause model ignorance of the common patterns, leading to sub-effective pattern learning~\cite{APG/NEURIPS2022_9cd0c571}. For APG, the AUC results indicate that it underfits the data in all scenarios, since low-rank decomposition makes its model capacity too low (even lower than an MLP), though it adds shared weights to the parameter generation process compared to M2M. In contrast, \model{} not only maintains high model capacity but also integrates shared FSLs. MuSeNet is basically the second best model, whose inferiority is probably due to ignoring the specific characteristics of multi-scenario route data.

\subsubsection{Ablation studies.}
To study the contribution of each component in \model{}, we first conduct the ablation study of the four proposed techniques on \dataset{} test. As shown in \cref{tab:ablation}, scenario factorization plays the most crucial role in model performance, followed by SAFF, $\mathcal{L}_{DR}$, and SABN. It is visible that they four all have a more significant improvement on $S_{12}^{36}$, $S_{37}^{61}$,$S_{62}^{72}$, than on the head $S_{1}^{11}$. Specifically, the drops of AUC  caused by removing SAFF or SABN validate their effectiveness, respectively.
Though removing $\mathcal{L}_{DR}$ only brings a drop of 0.2\% in the overall AUC, it results in a drop of at least 1.11\% in AUC$_{37}^{61}$ and AUC$_{62}^{72}$, which validates the effectiveness of $\mathcal{L}_{DR}$ in mining diverse factor scenarios. 

We further compare different versions of disentangling regularization in \cref{tab:ablation}, where $\mathcal{L}_{orth}=\Vert \bar{\mathbf{W}}^T\bar{\mathbf{W}}-I\Vert_F^2$ ($\bar{\mathbf{W}}$ is the concatenated matrix of $\{\bar{\mathbf{w}}^{(i)}\}_{i=1}^N$ and $\Vert \cdot\Vert_F$ is Frobenius norm) and $\mathcal{L}_{CNC}^{cos}$ denotes CNC in cosine form. 
Compared to \model{} without regularization, $\mathcal{L}_{orth}$ brings improvement to the head $S_{12}^{36}$, but is significantly worse than $\mathcal{L}_{DR}$ in the tail $S_{37}^{61}$, $S_{62}^{72}$, since $\mathcal{L}_{orth}$ tends to group $\bar{\mathbf{w}}^{(i)}$ closer~\cite{MMA20/NEURIPS2020_dcd2f3f3}. Against $\mathcal{L}_{orth}$, though $\mathcal{L}_{MMA}$ brings some gain in the tail $S_{37}^{61}$, $S_{62}^{72}$, due to angular diversity, the performance drops in the head scenarios as the gradient noise brought by $\mathcal{L}_{MMA}$. Against $\mathcal{L}_{MMA}$, $\mathcal{L}_{NCR}$ brings comprehensive improvement for addressing the gradient noise problem, but is still slightly behind $\mathcal{L}_{DR}$ since sufficient parameter disentangling cannot be achieved without $\mathcal{L}_{CNC}$. Due to gradient vanishing caused by the cosine similarity (\cref{eq:CNC_cos_grad}), $\mathcal{L}_{CNC}^{cos}$ works slightly worse in parameter disentangling, leading to the inferior performance. It is worth noting that $\mathcal{L}_{NCR}$ and $\mathcal{L}_{NCR}+\mathcal{L}_{CNC}^{cos}$ both bring better performance than $\mathcal{L}_{DR}$ in the tail $S_{62}^{72}$. We argue the reason is that the two regularizations have looser constraints on neuron directions clustering than $\mathcal{L}_{DR}$ and it results in neurons being more angularly separated, which brings more inducing effect on mining tail factor scenarios according to Assumption 3.1.


\begin{figure}[tbp]
  \centering
  \subfigure[Training curves of NCR \& CNC]{
    \includegraphics[width=0.476\linewidth]{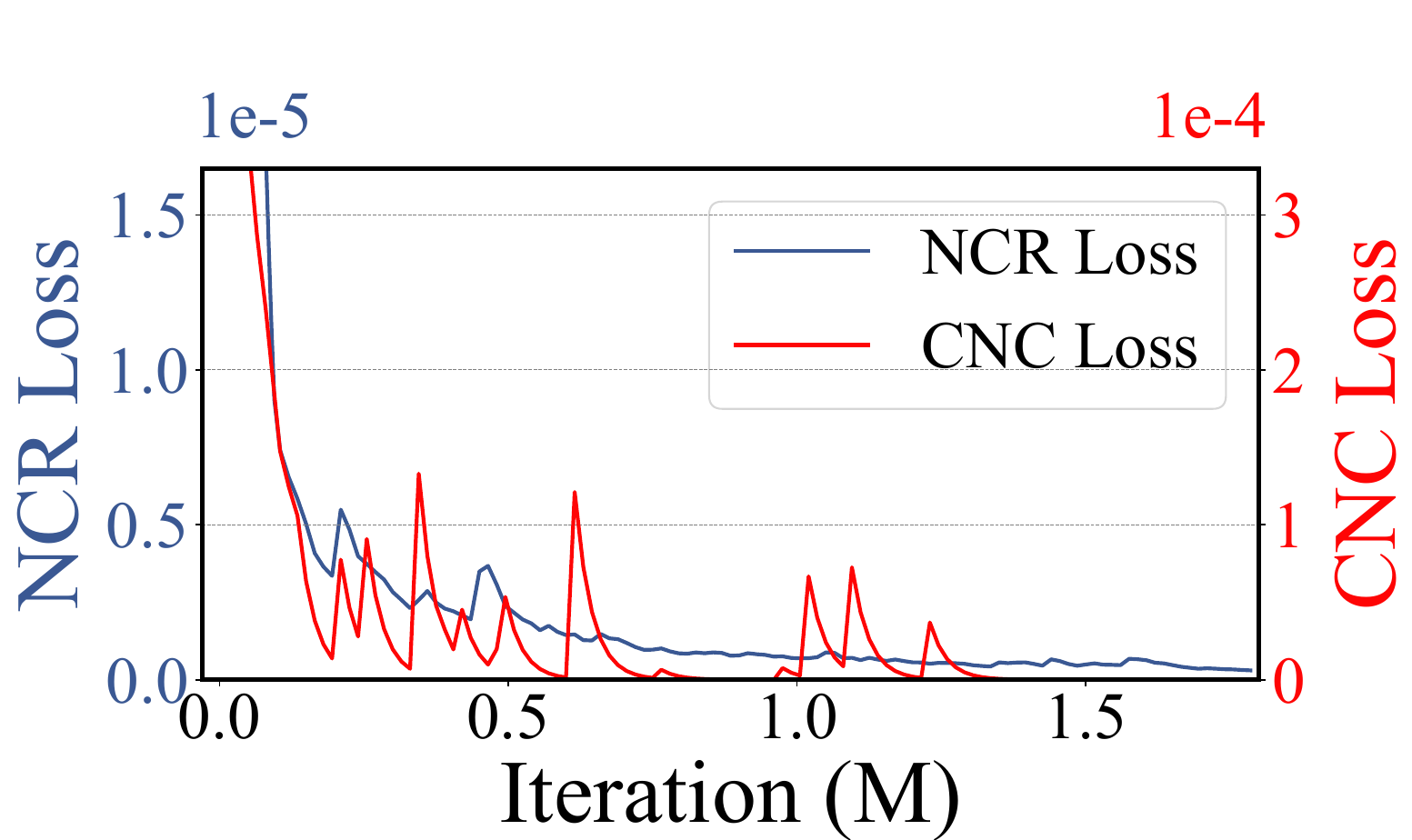}
    \label{fig:vis:a} }
   \hfill
  \subfigure[Sensitivity of $N$ \& $\kappa$ ]{
    \includegraphics[width=0.45\linewidth]{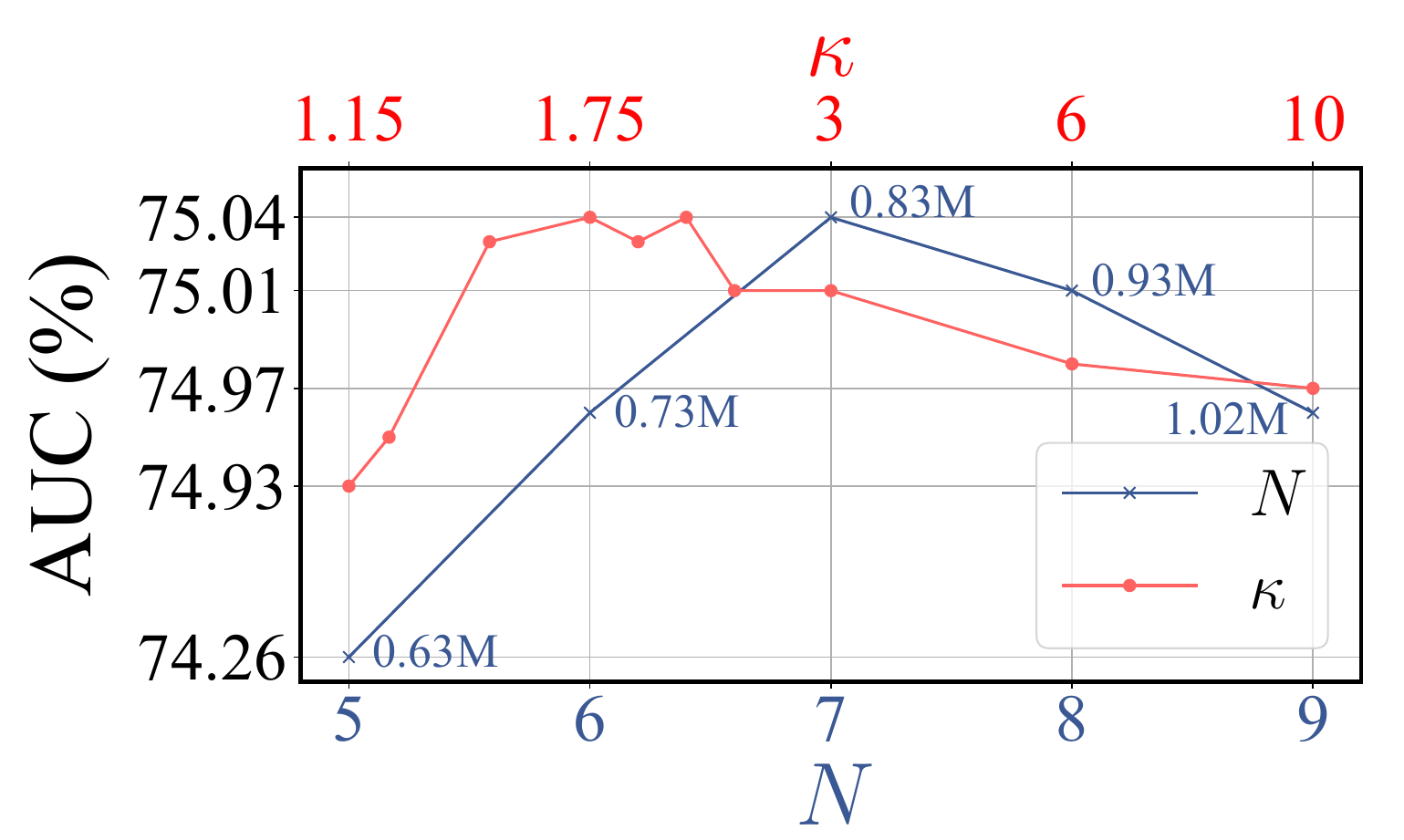}
    \label{fig:vis:b} }
  \caption{Visualization results. a) Convergence of our disentangling regularization $\mathcal{L}_{DR}$. b) Sensitivity of two key hyperparameters (\emph{i.e.} $N$ and $\delta_\theta$), where $\delta_\theta=\frac{1}{\kappa} \arccos(-\frac{1}{6})$ and $\kappa$ varies linearly within each pair of adjacent graduations.}
  \label{fig:vis}
\end{figure}

\subsubsection{Visualization.} \label{subsubsec:vis}
We further conduct experiments to analyze the convergence of our $\mathcal{L}_{DR}$ and the sensitivity of hyperparameters. As visible in \cref{fig:vis:a}, $\mathcal{L}_{NCR}$ and $\mathcal{L}_{CNC}$ both gradually converge during training, especially $\mathcal{L}_{CNC}$ converges to zero completely after oscillation, which not only validates the convergence of our $\mathcal{L}_{DR}$, but also implies that the angular separation among the FSLs is indeed realized in \model{}. We then present the impact of hyperparameters $N$ and $\delta_\theta$ on model performance in in \cref{fig:vis:b}, where we have $\delta_\theta=\frac{1}{\kappa} \arccos(-\frac{1}{6})$. For $N$, AUC increases with $N$ and peaks at $N = 7$ and then decreases, and the reason for the drop is possibly that larger $N$ makes data sparser for each factor scenario, thus reducing the overall performance. For ${\kappa}$, AUC is stable when ${\kappa}$ varies between 1.5 and 3.0. But AUC decreases rapidly when ${\kappa}$ is close to 1.0 since the threshold $\delta_\theta$ is close to $ \arccos(\frac{1}{1-N})$ where $N=7$, \emph{i.e.} the converged angle between neural centroids (Lemma 3.1), thus the threshold is too tight to maintain angular diversity, leading to the performance drop. Similarly, when ${\kappa}$ gets much larger, AUC gradually decreases since $\delta_\theta$ in CNC becomes too loose to induce sufficient disentanglement, leading to the performance drop.

\subsubsection{Interpretability.} \label{subsubsec:interpret}
\begin{figure}[tbp]

  \centering
   \includegraphics[width=1\linewidth]{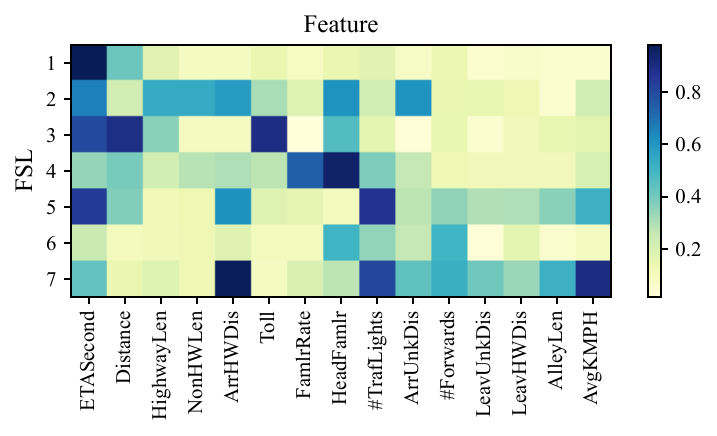}
   \caption{Interpretability experimental results, showing the route features each FSL focuses on. The explanations of the feature names are exhibited in \cref{tab:routefeat}.}
   \label{fig:interpret}
\end{figure}

Since \model{} actually builds an interpretable scenario space where each disentangled FSL represents an orthogonal coordinate basis, we conduct an interpretability study to unveil the disentangled factor scenario each FSL learns. Specifically, for each FSL $\mathbf{\tilde{\Theta}_i}$, we randomly sampled 200 data samples from \dataset{}, each of which had the corresponding gate value $\alpha_i$ greater than 0.8. We fed these data into DSFNet and modified the gates $\alpha$ to the one-hot vector which represents the FSL it belongs to, then obtained the logit $y$ before sigmoid. We use $\vert \frac{\partial y}{\partial \hat{\mathbf{x}}_k} \vert$ to describe the sensitivity of FSL to feature $\hat{\mathbf{x}}_k$, \emph{i.e.} the attention to $\hat{\mathbf{x}}_k$, and their values are normalized to [0,1] within each sample. Finally, we average the 200 gradient vectors to obtain the attention vector of the corresponding FSL. We select some significant route features and exhibit the corresponding attention values from all FSLs in \cref{fig:interpret}, which essentially reveals the diverse user preferences under different factor scenarios. As shown, the 1st FSL learns a common factor of considering time and distance, the 2nd one gives much focus on highway information, the 3rd reflects the concern of toll, the 4th depicts a preference to user-familiar routes, the 5th may learn a wait-hating preference, the 6th may learn to care about driving comfort, and the last probably learns a factor of preferring routes with high traffic flow speed. In addition, the diverse attentions further validate the disentanglement among the FSLs.

\subsubsection{Online A/B test.}
\model{} has been deployed in AMap to provide driving navigation services for non-holidays. Before deployment, we further trained \model{} offline on an extended version of \dataset{}, which included one month's data (excluding holidays) for all cities in China, then we conducted a one-week online A/B test.  The results of the online test are shown in \cref{tab:online}, which suggests the promotion of our \model{} brought to online industrial mapping applications on all metrics. Specifically, we adopt three online metrics, \emph{i.e.} deviation rate (DR), average number of deviations (AND), user coverage (UC), where DR indicates the probability of user’s deviation in a route, AND indicates the expectation of the number of user’s deviations in a navigation, and UC indicates the expectation of user's coverage rate of a route. Recall that AMap collects route label, selected routes set, ToD, and coverage rate of the route (\cref{subsec:data_acq}), hence we can obtain DR by calculating the proportion of routes labeled as negative in the selected routes in the database, AND by averaging ToDs in the database, and UC by averaging the coverage rates of routes from the selected routes in the database. 

\begin{table}[tb]
\scalebox{0.9}{
\begin{tabular}{@{}c|ccc|c@{}}
\toprule
\multirow{2}{*}{Metric} & \multicolumn{3}{c|}{ \textbf{DSFNet (Ours)}   \textit{vs.} Base} & \multicolumn{1}{c}{\textbf{DSFNet (Ours)}   \textit{vs.} Recall} \\ \cmidrule(l){2-5} 
                        & DR  $\downarrow$                 & AND $\downarrow$                & UC $\uparrow$                & DR $\downarrow$                                              \\ \midrule
$\mathbf{\Delta}$   & -1.61\%              & -0.86\%             & +3.82\%             & -6.21\%                                        \\ \bottomrule
\end{tabular}
}
\caption{The relative improvements on AMap in the online A/B test, where `Base' denotes the system with the latest previously deployed ranking model and `Recall' denotes the system without any ranking model (\emph{i.e.} only a recalling model).$\mathbf{\Delta}$ is relative improvement calculated by $(x-b)/b\times 100\%$ where $x$ and $b$ are metric values of \model{} and baseline, respectively. }
\label{tab:online}
\end{table}

\section{Conclusion}
In this paper, we specifically analyze the multi-scenario route ranking problem (MSRR) and propose \model{}, a novel method for MSRR that composes the dynamic parameters with a multi-branches structure where each branch represents a disentangled factor scenario. Specifically, we propose four novel techniques, \textit{i.e.} scenario factorization, disentangling regularization, scenario-aware batch normalization, and scenario-aware feature filtering, to arm a vanilla MLP into our \model{}. We demonstrate the superiority of \model{} via extensive experiments on the proposed large-scale industrial dataset \dataset{} and some online tests. In the future, we plan to study the applicability of \model{} in other fields, \emph{e.g.} e-commerce, to extend our method to multi-task ranking, and to explore how to automatically learn the number of factor scenarios in the training phase.


\appendix

\section{Proof of Lemma 3.1}\label{app:proof}
\begin{proof}
First, we prove the second `$\Leftrightarrow$' with $\forall_{i\neq j}\,\bm{\theta}^{(i,j)}=\arccos(\\-\frac{1}{N-1}) \Leftrightarrow \forall_{i\neq j}\, \Vert \bm{\theta}^{(i,j)}-\arccos(\frac{1}{1-N})\Vert^2=0 \Leftrightarrow \max_{i\neq j} \Vert \bm{\theta}^{(i,j)}-\arccos(\frac{1}{1-N})\Vert^2=0 \Leftrightarrow \min \max_{i\neq j} \Vert \bm{\theta}^{(i,j)}-\arccos(-\frac{1}{N-1})\Vert^2$. Then, we use $\Psi_1$, $\Psi_2$ to denote $\Psi_1: \min \max_{i\neq j} {\bar{\mathbf{w}}^{(i)}}\cdot\bar{\mathbf{w}}^{(j)}$ and $\Psi_2: \forall_{i\neq j}\,\bm{\theta}^{(i,j)}=\arccos(-\frac{1}{N-1})$, respectively, and we prove $\Psi_1\Leftrightarrow\Psi_2$ as follows. Since $\forall_i \,\Vert  \bar{\mathbf{w}}^{(i)} \Vert=1$, we have $N(N-1)\max_{i\neq j}\bar{\mathbf{w}}^{(i)}\cdot \bar{\mathbf{w}}^{(j)} 
     \geq \sum_{i\neq j} \bar{\mathbf{w}}^{(i)}\cdot \bar{\mathbf{w}}^{(j)} 
     = \Vert \sum_{i=1}^N \bar{\mathbf{w}}^{(i)} \Vert^2 - \sum_{i=1}^N \Vert  \bar{\mathbf{w}}^{(i)} \Vert^2 
     = \\\Vert \sum_{i=1}^N \bar{\mathbf{w}}^{(i)} \Vert^2 - N 
     \geq -N$, which indicates that $\min\max_{i\neq j} \bar{\mathbf{w}}^{(i)}\cdot \bar{\mathbf{w}}^{(j)}=\frac{1}{1-N}$ and the minimum is reached when $\forall_{k\neq l}\,\bar{\mathbf{w}}^{(k)}\cdot \bar{\mathbf{w}}^{(l)}=\frac{1}{1-N}$. Thus, we have $\Psi_2\Rightarrow\forall_{k\neq l}\,\bar{\mathbf{w}}^{(k)}\cdot \bar{\mathbf{w}}^{(l)}=\frac{1}{1-N}\Rightarrow\Psi_1$.
     But to prove $\Psi_1\Rightarrow\Psi_2$, we still need to prove that $\forall_{k\neq l}\,\bar{\mathbf{w}}^{(k)}\cdot \bar{\mathbf{w}}^{(l)}=\frac{1}{1-N}$ can be achieved in the $d$-dimensional space as $\bar{\mathbf{w}}^{(i)}\in\mathbb{R}^d$. It is equivalent to $\exists \bar{\mathbf{W}}\,\mathrm{s.t.}\, \bar{\mathbf{W}}^T\bar{\mathbf{W}}=\mathbf{P}$, where $\bar{\mathbf{W}}$ is the concatenated matrix of $\{ \bar{\mathbf{w}}^{(i)}\}_{i=1}^N$ and $\mathbf{P}$ is an $N \times N$ matrix where the diagonal is 1 and the remaining values are $\frac{1}{1-N}$. According to \cite{linearAlge}, $\mathbf{P}$ is a semi-positive definite matrix whose rank is $N-1$, thus $\bar{\mathbf{W}}$ can be obtained through the eigendecomposition of $\mathbf{P}$ when $d\geq N-1$, which proves $\Psi_1\Rightarrow\Psi_2$.
\end{proof}
\bibliographystyle{ACM-Reference-Format}
\bibliography{ref}

\end{document}